\documentclass[twocolumn,aps,pra,showpacs,superscriptaddress]{revtex4}
\usepackage{mathbbold}
\usepackage{amssymb}
\usepackage{graphicx}

\begin{document}

\title{Experimental high-intensity three-photon entangled source}
\author{Huai-Xin Lu}
\affiliation{Department of Physics and Electronic Science, Weifang University, Weifang,
Shandong 261061, China}
\author{Jun Zhang}
\thanks{Present address: Group of Applied Physics, University of Geneva, CH-1211 Geneva 4, Switzerland}
\affiliation{Hefei National Laboratory for Physical Sciences at
Microscale and Department of Modern Physics, University of Science
and Technology of China, Hefei, Anhui 230026, China}
\author{Xiao-Qin Wang}
\affiliation{Department of Physics and Electronic Science, Weifang University, Weifang,
Shandong 261061, China}
\author{Ying-De Li}
\affiliation{Department of Physics and Electronic Science, Weifang University, Weifang,
Shandong 261061, China}
\author{Chun-Yan Wang}
\affiliation{Department of Physics and Electronic Science, Weifang University, Weifang,
Shandong 261061, China}
\date{\today}

\begin{abstract}
We experimentally realize a high-intensity three-photon
Greenberger-Horne-Zeilinger (GHZ) entanglement source directly following the
proposal by Rarity and Tapster [J. G. Rarity and P. R. Tapster, Phys. Rev. A
\textbf{59}, R35 (1999)]. The threefold coincidence rate can be more than
$200$\,Hz with a fidelity of $0.811$, and the intensity can be further
improved with moderate fidelity degradation. The GHZ entanglement is
characterized by testing the Bell-Mermin inequality and using an
entanglement witness operator. To optimize the polarization-entangled
source, we theoretically analyze the relationship between the mean photon
number of the single-photon source and the probability of parametric
down-conversion.
\end{abstract}

\maketitle

\section{INTRODUCTION}

Recently optical quantum-information processing has been developed
rapidly both theoretically and experimentally, where photonic
entanglement, especially multiphoton entanglement~\cite{PCZWZ08} plays a crucial
role not only in the fundamental tests of quantum
nonlocality~\cite{B64}, but also in
optical quantum computation~\cite{KLM01,KMNRDM07} and quantum teleportation~%
\cite{Bennett93}, multiparty communication~\cite{HBB99}, quantum key
distribution~\cite{GRTZ02}. As we know, two-photon maximally entangled
state, or Bell state, can violate the Bell inequality~\cite{B64} to show the
inconsistency between quantum mechanics (QM) and local reality (LR) theory,
and so far there are numerous experiments to verify the validity of quantum
mechanics. However, entanglement of more than two photons, i.e.,
multiphoton maximally entangled state, or GHZ state~\cite{GHZ89,GHSZ90} can
demonstrate the conflicts deterministically and nonstatistically, which are
stronger and more straightforward than two-photon entanglement.

Several groups have experimentally realized the creation and manipulation of
multiphoton entanglement, from three-photon~\cite{BPDWZ99}, four-photon~%
\cite{PDGWZ01}, five-photon~\cite{ZCZYBP04} to six-photon~\cite{ZGWCZYMSP06}
entanglement. However, all of the reported multiphoton entangled sources so
far are hard to be applied to long-distance quantum communication due to the
limited intensity of the sources.

In this Letter, we report a high intensity of three-photon
polarization-entangled source following the proposal by Rarity and Tapster
\cite{RT99}. It is brighter than the intensities of publicly reported
three-photon entanglement in all the previous experiments so far. Although
in our experiment we use the same techniques as the previous experiments of
multiphoton entanglement, here we obtain brighter three-photon entanglement
with better quality through careful improvements and optimizations. Our
results suggest that the three-photon entangled source can be used for
multiparty long-distance quantum communication.


\section{EXPERIMENT}

\begin{figure}[tbp]
\includegraphics[scale=0.6]{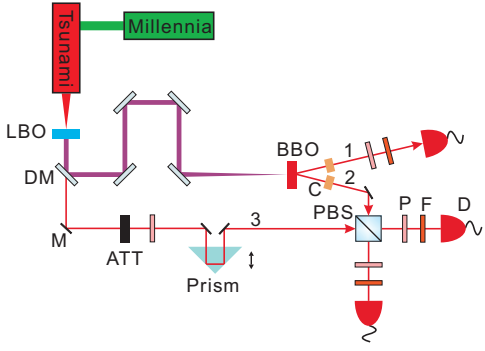}\newline
\caption{(Color online) Experimental setup. ATT, attenuator; M, mirror; C,
compensator; P, polarizer; F, filter; D, detector.}
\label{fig1}
\end{figure}

The experimental setup is shown in Fig.~\ref{fig1}. The cw $532$\,nm all
solid-state green laser (Millennia Pro 10s, Newport Co.) pumps the
mode-locked Ti:sapphire laser (Tsunami, Newport Co.) and generates
femtosecond pulses at the central wavelength of $780$\,nm, with a repetition
rate of $80$\,MHz and a pulse width of $\sim 100$\,fs. In the experiment,
the power of pump laser is $8.5$\,W and the output power of pulsed red laser
is $\sim 1.4$\,W. Then the pulsed laser passes through the LiB$_3$O$_5$
(LBO) crystal to produce ultraviolet (uv) laser at the central wavelength of
$390$\,nm after the up-conversion process. A lens is inserted before the LBO
crystal to form a small focused beam on the crystal and improve the
up-conversion efficiency. Since the output beam from the laser is
elliptical, we use the combination of two cylindrical lens to reshape the
beam to be circular. Due to the limited efficiency of up-conversion, the
beam after the LBO crystal is mixed by the uv and unconverted red laser. In
order to effectively separate them, five dichroic mirrors (DM), which can
reflect uv laser while transmitting red laser, are used. The power of the
transmitted unconverted $780$\,nm laser from the first DM is more than $500$%
\,mW. It is strongly attenuated by a series of attenuators to simulate a
single-photon source (SPS). The uv laser of $\sim 400$\,mW traverses a $2$%
\,-mm-thickness $\beta$-BaB$_2$O$_4$ (BBO) crystal to generate
polarization-entangled photon pairs owing to the process of spontaneous
parametric down-conversion~\cite{KMWZSS95}. Through optimizing the
collection efficiency, the observed count rate of photon pairs is more than $%
18$\,kHz, with a compensator composed of a half-wave-plate and a $1$%
\,-mm-thickness BBO crystal, and a $3$\,nm bandwidth interferometer filter
(IF) in each side. The prepared two-photon polarization-entangled state can
be written as
\begin{equation}  \label{2}
|\Psi\rangle_{12}=1/\sqrt{2}(|H\rangle_1|V\rangle_2 +
|V\rangle_1|H\rangle_2),
\end{equation}
where $|H\rangle$, $|V\rangle$ present horizontal and vertical polarization
of photons respectively. During the experiment, the visibility of two-photon
entangled state is $95\%$ in the H/V basis while $94\%$ in the +/- basis, with $%
|+\rangle = 1/\sqrt{2}(|H\rangle + |V\rangle)$ and $|-\rangle = 1/\sqrt{2}
(|H\rangle - |V\rangle)$.

The polarization of a single photon in path 3 is fixed at $|+\rangle$ and sent
to a polarization beam splitter (PBS), which transmits horizontal
polarization while reflecting vertical polarization. After carefully adjusting
the position of the prism controlled by a precise stepper motor, photon 2
and photon 3 can be completely overlapped on the PBS both spatially and
temporally. When the polarization of the two photons is identical, we can
prepare the three-photon GHZ entanglement as follows:
\begin{equation}  \label{3}
|\Psi\rangle_{123}=1/\sqrt{2}(|H\rangle_1|V\rangle_2V\rangle_3 +
|V\rangle_1|H\rangle_2H\rangle_3),
\end{equation}
since we can slightly tilt the BBO crystal in one of the compensators to
preserve that the phase of the state is zero.

\begin{figure}[tbp]
\centering \includegraphics[scale=0.65]{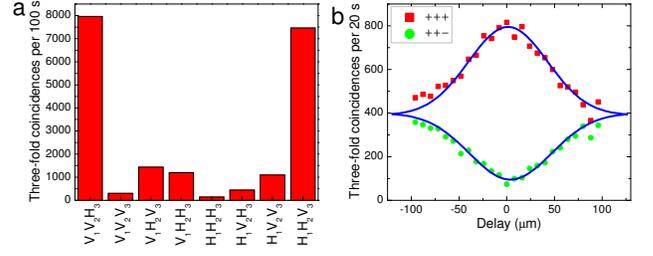}
\caption{(Color online) (a) Eight coincidence components in the H/V basis. (b)
Experimental observation of two-photon Hong-Ou-Mandel interference. Points
and solid lines present experimental values and theoretical fits. Threefold
coincidences in the +/- basis are measured as a function of the delay position
of the prism. At the zero delay, the interference visibility is more than
0.79, which is sufficient to imply the coherent superposition of three
photons.}
\label{fig2}
\end{figure}

The eight coincidence components are shown in Fig.~\ref{fig2}(a), where the
undesired components are much less than the desired coincidences of H$_1$V$%
_2 $V$_3$ and V$_1$H$_2$H$_3$. In order to confirm the coherent
superposition of the three-photon GHZ state and observe the two-photon
interference effect~\cite{HOM87}, we first move the position of the prism
and then perform the coincident measurement on the three photons in the +/-
basis, which is plotted in Fig.~\ref{fig2}(b).

\begin{figure}[tbp]
\centering \includegraphics[scale=0.45]{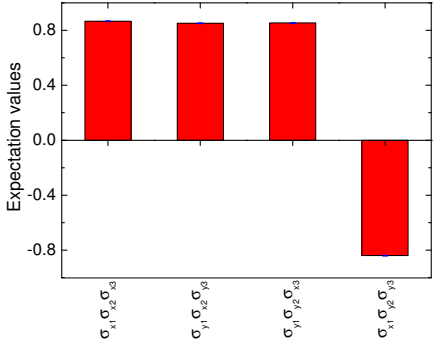}
\caption{(Color online) Experimental results of testing the Bell-Mermin
inequality.}
\label{fig3}
\end{figure}

We have emphasized that the GHZ state can explicitly show the conflicts
between QM and LR. Consider the following four joint measurements on the
three-photon GHZ state:
\begin{equation}  \label{JM}
\sigma_{x1}\sigma_{x2}\sigma_{x3}, \sigma_{x1}\sigma_{y2}\sigma_{y3},
\sigma_{y1}\sigma_{x2}\sigma_{y3}, \sigma_{y1}\sigma_{y2}\sigma_{x3},
\end{equation}
where $\sigma_{x}, \sigma_{y}$ are the Pauli operators, which can be
experimentally measured in the +/- and R/L basis, respectively, with $|R\rangle =
1/\sqrt{2}(|H\rangle + i|V\rangle)$ and $|L\rangle = 1/\sqrt{2}(|H\rangle -
i|V\rangle)$. Therefore, we can easily measure the above four joint
operators, e.g., $\sigma_{x1}\sigma_{y2}\sigma_{y3}$ implies that photon 1
is measured in the +/- basis while photon 2 and photon 3 are both measured in the
R/L basis simultaneously. Then we can experimentally test the Bell-Mermin
inequality~\cite{M90} by
\begin{equation}  \label{Mermin}
M=\sigma_{x1}\sigma_{x2}\sigma_{x3}+\sigma_{y1}\sigma_{x2}\sigma_{y3}+%
\sigma_{y1}\sigma_{y2}\sigma_{x3}-\sigma_{x1}\sigma_{y2}\sigma_{y3}.
\end{equation}
QM and LR will predict completely different results for this inequality with
\begin{equation}  \label{QMLR}
|<M>_{LR}|\le 2,\quad |<M>_{QM}|\le 4.
\end{equation}
That is to say, the expectation value of $M$ predicted by any LR theory
cannot be larger than 2. We verify this by performing polarization
measurements on the state. For example, to measure the expectation value of the $%
\sigma_{x1}\sigma_{y2}\sigma_{y3}$ operator, eight sorts of polarization
settings ($+RR, +RL, +LR, +LL, -RR, -RL, -LR$, and $-LL$) must be performed.
The result is shown in Fig.~\ref{fig3} and the experimental value of $M$ is
$M_E=3.4113\pm 0.0054$, which shows a violation of LR with more than 260
standard deviations. This also indicates that the prepared state in the
experiment is a genuine GHZ state.

In fact, except for using the Bell-Mermin inequality to validate the GHZ
entanglement, there are some other approaches to characterize the
multipartite entanglement such as quantum state tomography, entanglement
witness. We can quantify the quality of the produced GHZ state by evaluating
the fidelity as defined by
\begin{equation}  \label{fidelity}
\mathcal{F}=Tr (\rho_{exp}\vert\Psi_{123}\rangle\langle\Psi_{123}\vert),
\end{equation}
where $\rho_{exp}$ is the density matrix of the produced state. The fidelity
can be determined by the local measurements on individual qubits. The
experimental expectation value of fidelity is $\mathcal{F}_E=0.811\pm0.002$.

Furthermore, we can use the following entanglement witness operator~\cite%
{ABLS01} to detect the GHZ entanglement:
\begin{equation}  \label{witness}
\mathcal{W}=\frac{1}{2}\mathbbold{1}-|\Psi\rangle_{123}\langle\Psi|,
\end{equation}
where $\mathbbold{1}$ is an identity operator and $|\Psi\rangle_{123}\langle%
\Psi|$ is a projection operator to the GHZ state. If the expectation value
of this witness operator is negative, it implies the existence of genuine
GHZ entanglement. The expectation value of $\mathcal{W}$ is
\begin{equation}  \label{w}
\mathcal{W}_E=\frac{1}{2}-\mathcal{F}_E,
\end{equation}
whose experimental value is $\mathcal{W}_E=-0.311\pm0.002$, which is
negative with about 155 standard deviations. This clearly proves the
prepared state is truly entangled.



\section{EXPERIMENTAL ANALYSIS}

\begin{figure}[tbp]
\includegraphics[scale=0.65]{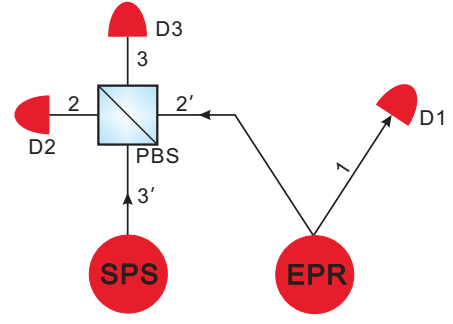}
\caption{(Color online) Simplified scheme for the three-photon entangled
source. SPS and EPR present the single-photon source and two-photon
entangled pairs respectively. }
\label{fig4}
\end{figure}

The single-photon source in path 3 is not a strictly true single-photon
emitter, but a weak coherent laser. The mean photon number $\mu $ of the SPS
will dominate the intensity of the three-photon entangled source. Higher $%
\mu $ will increase the intensity while decrease the visibility of the
entangled source. Thus optimization of the $\mu $ value is necessary. Here
we give a detailed analysis, see Fig.~\ref{fig4}. The total count rates
emitted from the SPS and Einstein-Podolsky-Rosen (EPR) pairs are
\begin{equation}
N_{S}=f\mu ,\qquad N_{E}=fP_{E},  \label{N}
\end{equation}%
where $f$ is the repetition frequency of the optical pulses, i.e., $80$%
\thinspace MHz, and $P_{E}$ is the probability of down-conversion per pulse
in the nonlinear crystal. For simplicity, we assume that the coupling
efficiencies of the three paths are equal, say, $\eta _{C}$, and the
detection efficiencies of the three detectors are also equal, say, $\eta _{D}
$, and the term of $e^{-\mu }$ is omitted. Then we can estimate the detected
count rates of $H_{3^{\prime }}$, $V_{3^{\prime }}$, $H_{3^{\prime
}}V_{3^{\prime }}$ from the SPS, and $H_{1}$, $V_{2^{\prime }}$, $V_{1}$, $%
H_{2^{\prime }}$, $H_{1}V_{2^{\prime }}$, $V_{1}H_{2^{\prime }}$ from the
EPR, respectively, given by
\begin{eqnarray}
&& N_{S(H_{3^{\prime }})}=N_{S(V_{3^{\prime }})}=\frac{1}{2}f\mu \eta
_{C}\eta _{D},  \nonumber  \label{SPSEPR} \\
&& N_{S(H_{3^{\prime }}V_{3^{\prime}})}=N_{S(V_{3^{\prime}}H_{3^{\prime }})}
=\frac{1}{2}f\mu ^{2}\eta_{C}^{2}\eta _{D}^{2}\times \frac{1}{4}=\frac{1}{8}f\mu ^{2}\eta
_{C}^{2}\eta _{D}^{2}, \nonumber \\
&& N_{E(H_{1})} =N_{E(V_{2^{\prime }})}=N_{E(V_{1})}=N_{E(H_{2^{\prime }})}=%
\frac{1}{2}fP_{E}\eta _{C}\eta _{D},  \nonumber \\
&& N_{E(H_{1}V_{2^{\prime }})}=N_{E(V_{1}H_{2^{\prime }})}=\frac{1}{2}%
fP_{E}\eta _{C}^{2}\eta _{D}^{2},
\end{eqnarray}%
where we assume that the PBS is ideal and the insertion losses of other
optical components are negligible. In the count rate of $N_{S(H_{3^{\prime
}}V_{3^{\prime }})}$, the factor of $\frac{1}{2}\mu ^{2}$ is due to the
consideration of Poisson distribution (the generation probability of two
photons from the SPS is $e^{-\mu }\mu ^{2}/2$), and the four sorts of
distributions of the two photons from the SPS ($H_{3^{\prime }}V_{3^{\prime
}},V_{3^{\prime }}H_{3^{\prime }},H_{3^{\prime }}H_{3^{\prime
}},V_{3^{\prime }}V_{3^{\prime }}$) result in the factor of $\frac{1}{4}$ in
$N_{S(H_{3^{\prime }}V_{3^{\prime }})}$.

When photon $2^{\prime }$ and photon $3^{\prime }$ are both $H$ or $V$
polarization, and the two photons are completely overlapped on the PBS, the
three-photon GHZ entangled state can be created and its count rate is
\begin{eqnarray}  \label{HVV}
N_{H_1V_2V_3} & = & N_{V_1H_2H_3}=\frac{N_{E(H_{1}V_{2^{\prime
}})}N_{S(H_{3^{\prime }})}}{f}  \nonumber \\
& = & \frac{1}{4}fP_E\mu\eta_C^3\eta_D^3.
\end{eqnarray}

Except the above two components, the other six sorts of the undesired
components will also contribute to the threefold coincidences. When $%
H_{3^{\prime }}$ photons are transmitted while $V_{3^{\prime }}$ photons are
reflected by the PBS, it will produce the following coincidence:
\begin{eqnarray}
N_{H_{1}V_{2}H_{3}} &=&N_{V_{1}V_{2}H_{3}}=\frac{N_{E(H_{1})}
\left [N_{S(H_{3^{\prime}}V_{3^{\prime }})}+N_{S(V_{3^{\prime}}H_{3^{\prime }})}\right ]}{f}  \nonumber  \label{HVH} \\
&=&\frac{1}{8}fP_{E}\mu ^{2}\eta _{C}^{3}\eta _{D}^{3}.
\end{eqnarray}%
Similarly, when $H_{2^{\prime }}$ photons and $V_{2^{\prime }}$ photons pass
the PBS it can cause the other two components
\begin{eqnarray}
N_{H_{1}H_{2}V_{3}} &=&N_{V_{1}H_{2}V_{3}}=\frac{N_{E(H_{1}V_{2^{\prime
}})}N_{E(H_{2^{\prime }})}}{f}  \nonumber  \label{HHV} \\
&=&\frac{1}{4}fP_{E}^{2}\eta _{C}^{3}\eta _{D}^{3}.
\end{eqnarray}%
The other undesired components of $H_{1}H_{2^{\prime }}$, $V_{1}V_{2^{\prime
}}$ can also cause the following coincidence
\begin{eqnarray}
N_{H_{1}H_{2}H_{3}} &=&N_{V_{1}V_{2}V_{3}}=\frac{N_{E(H_{1})}N_{E(H_{2^{%
\prime }})}N_{S(H_{3^{\prime }})}}{f^{2}}  \nonumber  \label{HHH} \\
&=&\frac{1}{8}fP_{E}^{2}\mu \eta _{C}^{3}\eta _{D}^{3}.
\end{eqnarray}

We can define a parameter $\mathcal{R}$ as the ratio of the desired
coincidences to the undesired coincidences
\begin{eqnarray}  \label{R}
\mathcal{R} & = & \frac{2N_{H_1V_2V_3}}{%
2(N_{H_1V_2H_3}+N_{H_1H_2V_3}+N_{H_1H_2H_3})}  \nonumber \\
& = & \frac{2\mu}{\mu^2+2P_E+P_E\mu}.
\end{eqnarray}
Obviously, when $\mathcal{R}$ is larger the visibility of the GHZ state will
be better. When $\mu\simeq \sqrt{2P_E}$
, $\mathcal{R}$ will reach the maximum value and the visibility of the
three-photon entangled state is the best. We experimentally verify this
theoretical result through controlling the $\mu$ value of the SPS. We find
out when $\mu_e=\mu\eta_C\eta_D\sim 0.014$ the visibility of the entangled
state can reach a quite good value. While during the experiment, we estimate
$P_E\sim 0.022$ and $\eta_C\eta_D$ in path 3 is $\sim 0.1$, therefore the
theoretical value of $\mu_e$ is $0.021$. The variance between the theoretical
and experimental values is mainly due to that the calibration for the coupling
efficiency in path 3 is not precise.

If $\mu$ is increased over the optimal value, the intensity of the
three-photon entangled source is also increased with the cost of the
decrease of the visibility. We observe the intensity to be more than $450$%
\,Hz with a few percentages of visibility degradation, which is larger than
the intensity in the first three-photon experiment~\cite{BPDWZ99} with more
than four orders of magnitude. On the other hand, in the experiment we have
tried to improve the pump of the laser system up to the maximum power, $\sim
10.5$\,W. Through careful optimization, the uv power can be increased to $%
\sim 500$\,mW and the intensity of two-photon entangled pairs is close to $%
30 $\,kHz, which also implies that we can boost the intensity of the
three-photon entangled source up to $\sim 1$\,kHz with a moderate
visibility. However, the laser system cannot be stable with this pump power
for long-term experiment. This high intensity of three-photon entangled
source is very valuable for many multiparty quantum communication protocols~%
\cite{HBB99} and other applications in quantum-information processing~\cite%
{BEZ00}.


\section{CONCLUSION}

In summary, we have experimentally implemented a bright three-photon
polarization-entangled source and tested the Bell-Mermin inequality to show
the conflicts between quantum mechanics and local realism. We also introduce
an entanglement witness operator to characterize three-photon GHZ
entanglement. Further, we analyze the theory of visibility optimization,
which is proven by the experimental results. The implementation of a high
intensity of three-photon entangled source is a significant step towards
practical long-distance multiparty quantum communication in the future.


\begin{acknowledgments}
This work was supported by the Natural Science Foundation
of China (Grant No. 60878001) and by the Natural
Science Foundation of Shandong Province (Grant No. Y2006A24).
\end{acknowledgments}



\end{document}